%% file: Energy-Efficient_RRM_Extended.tex
\documentclass[12pt, draftclsnofoot, onecolumn]{IEEEtran}	% regular single-column
 \usepackage{amsmath}
 \usepackage{amssymb}
 \usepackage{subfigure}
 \usepackage{graphicx}
 \usepackage{graphics}
 \usepackage{color}
 \usepackage{psfrag}
 \usepackage{cite}
 \usepackage{balance}
 \usepackage{algorithm}
 \usepackage{accents}
 \usepackage{amsthm}
 \usepackage{bm}
 \usepackage{url}
 \usepackage{algorithmic}
 \usepackage[english]{babel}
 \usepackage{multirow}
 \usepackage{enumerate}
 \usepackage{cases}
 \usepackage{stfloats}
 \usepackage{dsfont}
 \usepackage{soul}
 \usepackage{amsfonts}
 \usepackage{fancyhdr}
 \usepackage{hhline}
 \usepackage{array}
 \usepackage{booktabs}
 \usepackage{framed}
 \usepackage{float}

\include{header}

\begin{document}

\title{\huge Energy-Efficient Radio Resource Allocation for Federated Edge Learning}

\author{Qunsong Zeng, Yuqing Du, Kin K. Leung, and Kaibin Huang \\ 
\thanks{Q. Zeng, Y. Du and K. Huang are with  The University of Hong Kong, Hong Kong (Email: qszeng@eee.hku.hk, yqdu@eee.hku.hk, huangkb@eee.hku.hk). K. K. Leung is with Imperial College London, UK (Email: kin.leung@imperial.ac.uk).}
}

\maketitle
%============================================================
\begin{abstract}
Edge machine learning involves the development of learning algorithms at the network edge to leverage massive distributed data and computation resources. Among others, the framework of \emph{federated edge learning} (FEEL) is particularly promising for its data-privacy preservation. FEEL coordinates global model training at a server and local model training at edge devices over wireless links. In this work, we explore the new direction of energy-efficient \emph{radio resource management} (RRM) for FEEL. To reduce devices' energy consumption, we propose energy-efficient strategies for bandwidth allocation and scheduling. They adapt to devices' channel states and computation capacities so as to reduce their sum energy consumption while warranting learning performance. In contrast with the traditional rate-maximization designs, the derived optimal policies allocate more bandwidth to those scheduled devices with weaker channels or poorer computation capacities, which are the bottlenecks of synchronized model updates in FEEL. On the other hand, the scheduling priority function derived in closed form gives preferences to devices with better channels and computation capacities. Substantial energy reduction contributed by the proposed strategies is demonstrated in learning experiments.
\end{abstract}
%============================================================
\section{introduction}
Recent years have witnessed a phenomenal growth in mobile data, most of which are generated in real-time and distributed at edge devices (e.g., smartphones and sensors)~\cite{gxzhu_2018_edge_learning}. Uploading these massive data to the cloud for training \emph{artificial intelligence} (AI) models is impractical due to various issues including privacy, network congestion, and latency. To address these issues, the \emph{federated edge learning} (FEEL) framework has been developed~\cite{gxzhu2018FEEL, deniz2019federated_edge_learning, sqwang_2018_edge_learning}, which implements distributed machine learning at the network edge. In particular, a server updates a global model by aggregating local models (or stochastic gradients) transmitted by devices that are computed using local datasets. The updating of the global model using local models and the reverse are iterated till they converge. Besides preserving data privacy by avoiding data uploading, FEEL leverages distributed computation resources as well as allows rapid access to the real-time data generated by edge devices. One focus in the research area is communication-efficient FEEL where wireless techniques are designed to accelerate learning by reducing communication overhead and latency. However, the topic of energy-efficient communication for FEEL so far has not been explored. This is an important topic as training and transmission of large-scale models are energy consuming, while most edge devices especially sensors have limited battery lives. This topic is investigated in the current work where novel \emph{radio-resource-management} (RRM) strategies for joint bandwidth allocation and scheduling are proposed for minimizing the total device energy-consumption under a constraint on the learning speed. 

The topic of communication-efficient FEEL has been extensively studied from different aspects. One branch of research focuses on edge-device selection so as to accelerate learning~\cite{nishio2018background,yang2018background}. In particular, a partial averaging scheme is proposed in~\cite{nishio2018background}, where only a portion of updates from fast-responding devices are used for global updating while those from stragglers are discarded. However, perfect device-update-uploading is assumed, which ignores the hostility of wireless channels, and at the same time overlooks the possibility of exploiting the sophisticated properties of wireless channels for improving the communication efficiency. By taking the properties into account, a joint device-selection and beamforming design is proposed for accelerating the federated edge learning~\cite{yang2018background}. Nevertheless, the device selection criterion is only based on the \emph{channel-state information} (CSI) while ignoring the heterogeneous computation capacities of devices. On the other hand, to overcome the multi-access bottleneck, a \emph{broadband analog aggregation} (BAA) multiple-access scheme is proposed in~\cite{gxzhu2018FEEL}. Specifically, by exploiting the waveform-superposition property of a multi-access channel, updates simultaneously transmitted by devices over broadband channels are analog aggregated ``over-the-air" so as to reduce the multi-access latency. Given fixed communication cost per uploading, a control algorithm on uploading frequency is proposed in~\cite{sqwang_2018_edge_learning} by analyzing the convergence bound of distributed gradient descent to improve the learning performance. However, all these existing schemes are designed from the learning perspective while the energy-consumption issue of edge devices is out of scope, which is becoming increasingly important given the limited battery lives of devices. This motivates the current work on energy-efficient FEEL.

In this work, we consider the problem of minimizing energy consumption of edge devices in the context of FEEL without compromising learning performance. To this end, two energy-efficient RRM strategies are proposed for joint bandwidth allocation and scheduling. To the best known of authors' knowledge, this work represents the first attempt to consider the energy-efficient RRM for FEEL. 

To design the first energy-efficient RRM strategy, we assume a given set of edge devices and focus on bandwidth allocation. The optimal policy for energy minimization is derived in closed-form. The solution suggests that each edge device should utilize all the allowed uploading time so as to minimize the energy consumption. Furthermore, it can be observed from the solution that under the constraint of synchronous updates, less bandwidth should be allocated to devices with more powerful computation capacities and better channel conditions. This is in contrast with the traditional rate-maximization design.

The second strategy extends the first to include scheduling, namely selecting devices to participate in FEEL. We propose a practical algorithm for iterating between solving two sub-problems under the criterion of energy minimization: 1) scheduling and 2) bandwidth allocation using the first strategy. For scheduling, the optimal policy is derived in closed-form, indicating the selection priorities for devices. The solution suggests that a device with a poor computation capacity and a bad channel has a lower priority to be selected and vice versa. 

The remainder of this paper is organized as follows. Section~\ref{system model} introduces the system model. Sections~\ref{Section:Energy-efficient RRM} and~\ref{Section:joint RRM-selection} present two energy-efficient RRM strategies. Simulation results are provided in Section~\ref{Simulation}, followed by the concluding remarks in Section~\ref{conclusion}.

\begin{figure}[t]
    \centering
    \includegraphics[width=1.0\textwidth]{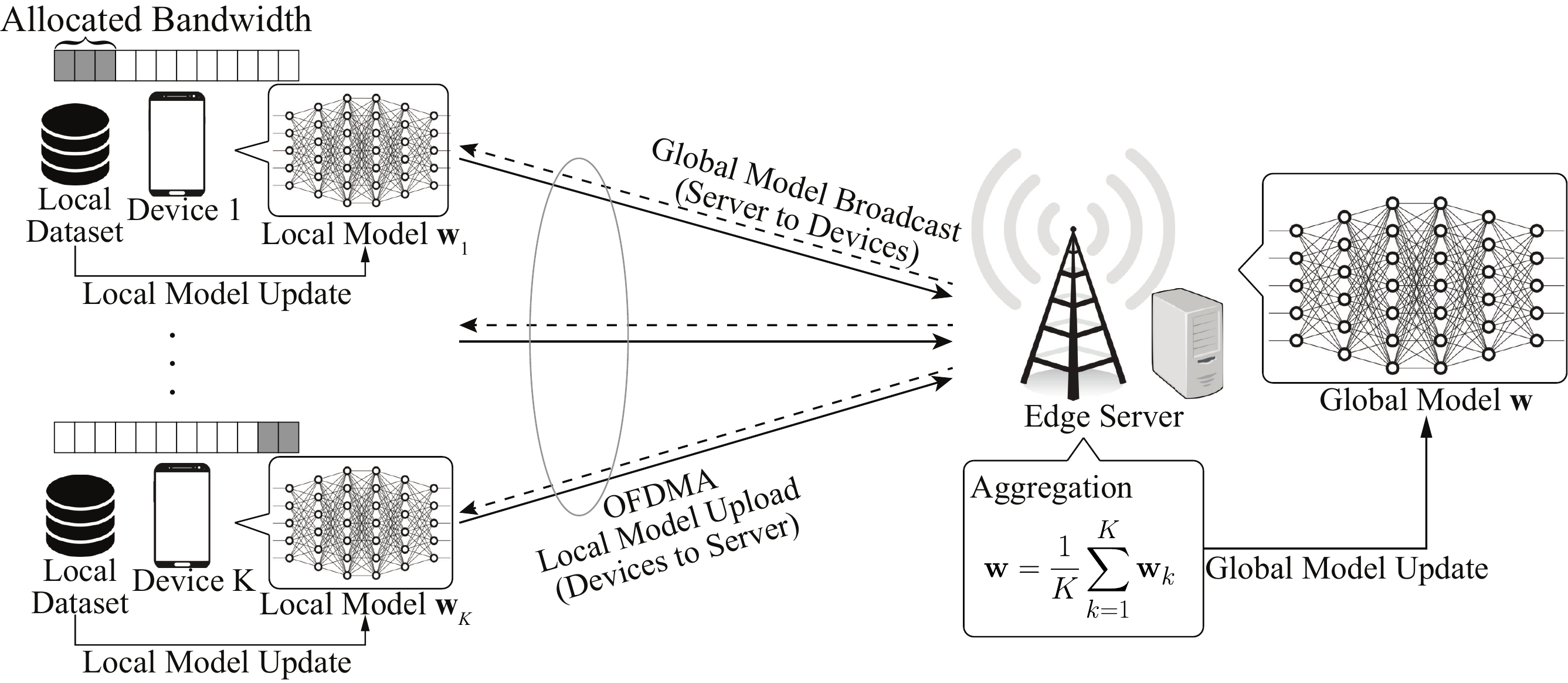}
    \vspace{-12mm}
    \caption{A framework for FEEL system.}
    \vspace{-10mm}
      \label{Fig:system_model}
\end{figure}
%============================================================
\section{System Model}\label{system model}
Consider a FEEL system consisting of a single edge server and $K$ edge devices, denoted by a set $\mathcal{K}=\{1,\cdots,K\}$. As described earlier and illustrated in Fig.~\ref{Fig:system_model}, FEEL iterates between two steps: 1) updating the global model at the server by aggregating local models transmitted over a multi-access channel; 2) replacing the local models by broadcasting the global model. Each iteration is called a \emph{communication round}. It is assumed that the edge server has perfect knowledge of the model size (determining the sizes of data transmitted by devices) as well as multiuser channel gains and local computation capacities, which can be obtained by feedback. Using this information, for each communication round, the edge server needs to determine the energy-efficient strategy for scheduling and allocating bandwidth. Due to the fact that communication rounds are independent, it suffices to consider the problem for an arbitrary round without loss of generality.

%------------------------------------------------------------
\subsection{Multiple-access Model}
Consider \emph{orthogonal frequency-division multiple access} (OFDMA) for local model uploading with total bandwidth $B$. Define $\gamma_k\in[0,1]$ as the bandwidth allocation ratio for device $k$, and the resulting allocated bandwidth is $\gamma_kB$. Furthermore, let $h_k$ denote the corresponding channel gain. Given synchronous updates~\cite{federatedlearning}, a time constraint is set for local model training and model uploading in each communication round as follows
\begin{equation}\label{InEq: time_constraint}
\text{(Time constraint)}\quad t_k^{\text{comp}}+t_k^{\text{up}}\leq T,\quad \forall k\in\mathcal K,
\end{equation}
where $t_k^{\text{comp}}$ and $t_k^{\text{up}}$ denote the time for local model training time and model uploading time of device $k$, respectively. $T$ is the maximum total time. The fact that edge devices have heterogeneous computation capacity is reflected in the differences among the values of $\{t_k^{\text{comp}}\}$. For ease of notation, we re-denote $t_k^{\text{up}}$ as $t_k$ hereafter. Then, it follows that~\eqref{InEq: time_constraint} can be rewritten as 
\begin{equation}
t_k\leq T_k,~\forall k\in\mathcal{K},
\end{equation}
where $T_k =T-t_k^{\text{comp}}$ is referred to as the allowed time for model uploading.

%------------------------------------------------------------
\subsection{Energy Consumption Model}
For each communication round, the energy consumption of a typical edge device comprises two parts: one for transmission (model uploading) and the other for local model training, which are specified in the following.
\subsubsection{Energy consumption for model uploading} Let $p_k$ denote the transmission power (in Watt/Hz) of device $k$. The achievable rate (in bit/s), denoted by $r_k$, can be written as
\begin{align}\label{Eq:shannon}
    r_k=\gamma_kB\log\left(1+\frac{p_kh_k^2}{N_0}\right),
\end{align}
where $N_0$ is the variance of the complex white Gaussian channel noise. Let $L$ denote the data size (in bit), the data rate can then be calculated as
\begin{equation}\label{Eq:data_rate}
 r_k = \frac{\beta_kL}{t_k},
\end{equation}
where $\beta_k$ is a \emph{state indicator} for device $k$. Specifically, $\beta_k = 1$ if device $k$ is selected for uploading, or $0$ otherwise. By combining~\eqref{Eq:shannon} and~\eqref{Eq:data_rate}, the uploading energy consumption can be calculated as 
\begin{equation}\label{Eq:E_upload}
E_k^{\text{up}}=\gamma_kBp_kt_k = \frac{\gamma_kBt_kN_0}{h_k^2}\left(2^{\frac{\beta_kL}{\gamma_kBt_k}}-1\right).
\end{equation}
\subsubsection{Energy consumption for local training}
Consider the local training of a \emph{neural network} model via the well-known \emph{backpropagation} (BP) algorithm on GPU. According to experiments reported in~\cite{mei2017GPUenergy}, the energy consumption of GPU only depends on the complexity of the BP algorithm and the size (or equivalently dimensions) of the model parameters. Since all edge devices train the same model of size $L$ using the BP algorithm, the energy consumption of edge devices for local training is identical and denoted as $E^{\text{comp}}$.

%----------------------------------------------------------
\subsection{Learning Speed Model}
It is proved in~\cite{bernstein2018sgd,yang2018background} that the convergence rate of distributed SGD is \emph{inversely proportional} to the number of participating devices. Therefore, we use the total number of scheduled devices as the measurement of learning speed. By leveraging the indicator $\{\beta_k\}$, the learning speed can be expressed as 
\begin{equation} 
\text{(Learning speed)}\quad\sum_{k=1}^K\beta_k.
\end{equation}
From the perspective of accelerating learning, it is desirable for the server to schedule as many devices as possible, which, however, is limited by finite radio resources.

%============================================================
\section{Energy-Efficient Bandwidth Allocation}\label{Section:Energy-efficient RRM}

In this section, we consider the problem of RRM for a given set of active devices which can all meet the time constraint in~\eqref{InEq: time_constraint} with $\beta_k = 1,\forall k\in \mathcal{K}$. The goal is to minimize the total energy consumption, i.e. $\sum_{k=1}^K\left(E_k^{\text{comp}}+E_k^{\text{up}}\right)$. Since the energy consumption for local model training, i.e. $E^{\text{comp}}$, is uniform and fixed, the problem focuses on minimizing uploading energy and thus is formulated as
\begin{equation}\label{opt_rrm}({\bf P1})\quad
\begin{aligned}
    \min_{\{\gamma_k,t_k\}}&~\sum_{k=1}^K\frac{\gamma_kBt_kN_0}{h_k^2}\left(2^{\frac{\beta_kL}{\gamma_kBt_k}}-1\right)\\
    \text{s.t. }&~\sum_{k=1}^K\gamma_k=1,~~0\leq \gamma_k\leq 1,~k\in\mathcal{K},\\
    &~~0\leq t_k\leq T_k,~k\in\mathcal{K}.\nn
\end{aligned}
\end{equation}
By solving the above problem, the server can optimally determine bandwidth partitioning, as specified by $\{\gamma_k\}$, and the uploading time $\{t_k\}$ for devices. To begin with, one basic characteristic of Problem $({\bf P1})$ is given as follows.

\begin{lemma}\label{lemma_non_increasing}
\emph{The objective of Problem $({\bf P1})$ is a non-increasing function in $t_k$ and $\gamma_k,\forall k\in\mathcal{K}$.}
\end{lemma}
The result follows from observing the derivative of the objective with the details omitted for brevity. It can be inferred from the Lemma~\ref{lemma_non_increasing} that it is optimal to maximize the transmission time of each device, resulting in $t^{\star}_k = T_k,\forall k\in \mathcal{K}$, which is independent of the allocated bandwidth $\gamma_k$. Then, it follows that the optimal RRM policy is obtained as follows.

\begin{theorem}\label{thm_opt_rrm}
\emph{(Optimal Bandwidth Allocation).
The optimal policy for bandwidth allocation is
\begin{align}
    \gamma_k^{\star}&=\frac{\beta_kL\ln{2}}{BT_k\left[1+\mathcal{W}\left(\frac{h_k^2\nu^{\star}-BT_kN_0}{BT_kN_0e}\right)\right]}, \quad \forall k\in\mathcal{K}\label{gamma_star},\\
    t_k^{\star}&=T_k,\quad \forall k\in\mathcal{K}\label{t_star},
\end{align}
where $\mathcal{W}(\cdot)$ is the Lambert $W$ function, $T_k=T-t_k^{\text{comp}}$ is the restricted transmission time for device $k$, $\nu^{\star}$ is the solved value for the Lagrange multiplier and $e$ is the Euler's number.
\begin{proof}
See Appendix \ref{proof_opt_tk}.
\end{proof}}
\end{theorem}
Next, to gain more insight, a corollary is given as follows.

\begin{corollary}\label{gamma_Tk}
\normalfont 
$\gamma_k^{\star}$ is a non-increasing function with respect to $T_k$ and $h_k^2$, respectively.
\begin{proof}
See Appendix \ref{proof_gamma_Tk}.
\end{proof}
\end{corollary}

One observation can be made from Corollary \ref{gamma_Tk} is that more bandwidths should be allocated to edge devices with weaker computation capacities, namely smaller $T_k$. The reason is that these devices are the bottlenecks in synchronized updates and sum energy minimization. To be specific, they require larger bandwidths so as to complete model uploading within the short allowed transmission/uploading time and also to reduce transmission power.

Furthermore, it can be observed that more bandwidths should be allocated to devices with weaker channels. Overcoming the conditions requires boosting transmission power or more bandwidths. For energy minimization, the latter is preferred. 

\begin{remark}
\emph{(Rate-centric vs. Learning-centric RRM).
The conventional RRM strategies for sum-rate maximization, such as \emph{water-filling},  allocate more resources to users with stronger channels. In contrast, the proposed RRM policy for edge learning allocates more resources to users with weaker channels and/or poorer computation capacities. This reflects the differences in communication principles for the two paradigms of communication-computation separation and communication-computation integration.}
\end{remark}

%============================================================
\section{Energy-and-Learning Aware Scheduling}\label{Section:joint RRM-selection}

In the presence of devices with poor computation capacities or weak channels, scheduling only a subset of devices for model uploading can reduce sum energy consumption as well as meet the time constraint. By modifying Problem $({\bf P1})$ to include the learning speed in the objective, the current problem can be formulated as
\begin{equation}\label{opt_joint}({\bf P2})\quad
\begin{aligned}
    \min_{\{\gamma_k,t_k,\beta_k\}}&~\sum_{k=1}^K\frac{\gamma_kBt_kN_0}{h_k^2}\left(2^{\frac{\beta_kL}{\gamma_kBt_k}}-1\right)-\lambda\sum_{k=1}^K\beta_k\\
    \text{s.t. }&~\beta_k\in\{0,1\},~k\in\mathcal{K},\\
    &~\sum_{k=1}^K\gamma_k=1,~~0\leq \gamma_k\leq 1,~k\in\mathcal{K},\\
    &~~0\leq t_k\leq T_k,~k\in\mathcal{K},\nn
\end{aligned}
\end{equation}
where the trade-off factor $\lambda>0$ is a pre-determined constant. Directly solving the above problem is difficult due to its non-convexity arising from the integer constraint. To solve this problem, we adopt the common solution method, referred to as \emph{relaxation-and-rounding}. Specifically, it firstly relaxes the integer constraint $\beta_k\in\{0,1\}$ as the real-value constraint $0\leq\beta_k\leq 1$, and then the integer solution is determined using rounding techniques after solving the relaxed problem. It is also noted that after this relaxation, the continuous value of $\beta_k$ can be viewed as the selection priority of edge device $k$. Mathematically, the relaxed problem can be written as
\begin{equation}\label{opt_joint_relaxed}({\bf P3})\quad
\begin{aligned}
    \min_{\{\gamma_k,t_k,\beta_k\}}&~\sum_{k=1}^K\frac{\gamma_kBt_kN_0}{h_k^2}\left(2^{\frac{\beta_kL}{\gamma_kBt_k}}-1\right)-\lambda\sum_{k=1}^K\beta_k\\
    \text{s.t. }&~0\leq\beta_k\leq 1,~k\in\mathcal{K},\\
    &~\sum_{k=1}^K\gamma_k=1,~~0\leq \gamma_k\leq 1,~k\in\mathcal{K},\\
    &~~0\leq t_k\leq T_k,~k\in\mathcal{K}.\nn
\end{aligned}
\end{equation}
It is easy to prove that $({\bf P3})$ is a convex problem. A standard solution approach is to use a numerical method since the optimization variables are all coupled. In the remainder of the section, we propose a more insightful and efficient approach that iterates between solving two sub-problems: 1) the bandwidth-allocation in Problem $({\bf P1})$; 2) scheduling problem. To be specific, the first sub-problem is to allocate bandwidths given scheduled devices indicated by $\{\beta_k\}$, where the optimal solution is given in Theorem~\ref{thm_opt_rrm}. The other sub-problem (scheduling) is to decide the selection priorities of edge devices, i.e. $\{\beta_k\}$, given $\{\gamma_k,t_k\}$, which can be mathematically written as

\begin{equation}\label{opt_beta}({\bf P4})\quad
\begin{aligned}
    \min_{\{\beta_k\}}&~\sum_{k=1}^K\frac{\gamma_kBt_kN_0}{h_k^2}\left(2^{\frac{\beta_kL}{\gamma_kBt_k}}-1\right)-\lambda\sum_{k=1}^K\beta_k\\
    \text{s.t. }&~~0\leq\beta_k\leq 1,~k\in\mathcal{K}.\nn
\end{aligned}
\end{equation}
It is easy to show that Problem $({\bf P4})$ is convex and the closed-form solution is derived in the following  Theorem.

%------------------------------------------------------------

\begin{theorem}\emph{(Edge-device Selection Priority).\label{thm_opt_user}
\normalfont
The optimal selection priority for device $k$ is given as 
\begin{align}\label{beta_star}
    \beta_k^{\star}=\min\left\{\max\left\{\frac{\gamma_kBT_k}{L}\log\left({\frac{\lambda h_k^2}{N_0L\ln{2}}}\right), 0\right\}1\right\},\quad k\in\mathcal{K}.
\end{align}
\begin{proof}
See Appendix \ref{proof_opt_betak}.
\end{proof}}
\end{theorem}

This theorem is  consistent with the intuition that device $k$ with a high computation capacity and a good channel should have a high priority to be selected, i.e. $\beta_k^{\star}$ is large.

\begin{remark}
\emph{(Effects of Parameters on Selection Priority).
It can be observed from~\eqref{beta_star} that $\beta_k$, indicating the selection priority of device $k$, scales with the allowed transmission time, i.e. $T_k$, linearly and with the channel gain approximately as $\log(h_k)$. The former scaling is much faster than the latter. This shows that the allowed transmission time (or equivalently computation capacity) is dominant over the channel on determining the selection priority of the device.}
\end{remark}
Based on the above results, the solution of Problem $({\bf P2})$ is provided in Algorithm \ref{Joint RRM} by iteratively solving $({\bf P1})$ and $({\bf P4})$ until convergence.
% Based on the above results, the solution of Problem (P2) is provided in Algorithm \ref{Joint RRM}.
\setlength{\textfloatsep}{4pt}
\begin{algorithm} [t!]
    \caption{Joint Bandwidth Allocation and Scheduling}
    \label{Joint RRM}
    \textbf{Initialization}: Randomly set indicators $\{\beta_k\}\in[0,1]$.\\
    \textbf{Iteration:}
    \begin{itemize}
    \item \textbf{(Energy-efficient Bandwidth Allocation)}: Given fixed $\{\beta_k\}$, compute $\{\gamma_k,t_k\}$ using (\ref{gamma_star}) and (\ref{t_star});    
    \item \textbf{(Energy-and-Learning Aware Scheduling)}: Given fixed $\{\gamma_k,t_k\}$, compute $\{\beta_k\}$ using (\ref{beta_star});   
    \end{itemize}
    \textbf{Until Convergence}.\\
    \textbf{Round} indicators $\{\beta_k\}$ to $\{0,1\}$.\\
    \textbf{Compute} $\{\gamma_k,t_k\}$ using (\ref{gamma_star}) and (\ref{t_star}).\\
    \textbf{Output} the optimal solution $\{\beta_k^{\star},\gamma_k^{\star},t_k^{\star}\}$.
\end{algorithm}

%============================================================
\section{Simulation results}\label{Simulation}
The simulation settings are as follows unless specified otherwise. There are $K=50$ edge devices with local model training time, $\{t^{\text{comp}}_k\}$, following the uniform distribution in the range of $(0, 10]$ ms. Consider an OFDMA system where the bandwidth $B=1$ MHz. The channel gains $\{h_k\}$ are modeled as independent Rayleigh fading with average path loss set as $10^{-4}$. The variance of the complex white Gaussian channel noise is set as $N_0=10^{-8}$ W. For learning, the model size is set to be $L=10^{4}$ bits and the task aims at classifying handwritten digits using the MNIST dataset. Each device is randomly assigned $20$ samples. The model is a 6-layer \emph{convolutional neural network} (CNN), consisting of two $5\times 5$ convolution layers with \emph{rectified linear unit} (ReLU) activation, which have $10$ and $20$ channels respectively, each followed by $2\times 2$ max pooling, a fully connected layer with $50$ units and ReLU activation, and a softmax output layer. 

% \begin{figure}[t]
%     \centering
%     \includegraphics[width=0.7\textwidth]{Figures/energy.eps}
%     \vspace{-5mm}
%     \caption{Sum device energy consumption vs. constrained time per communication round in a FEEL system.}
%     \vspace{-10mm}
%       \label{Fig:energy_consumption}
% \end{figure}

\subsubsection{Energy-efficient bandwidth allocation} 
\begin{figure}[t!]
    \centering
    \includegraphics[width=0.7\textwidth]{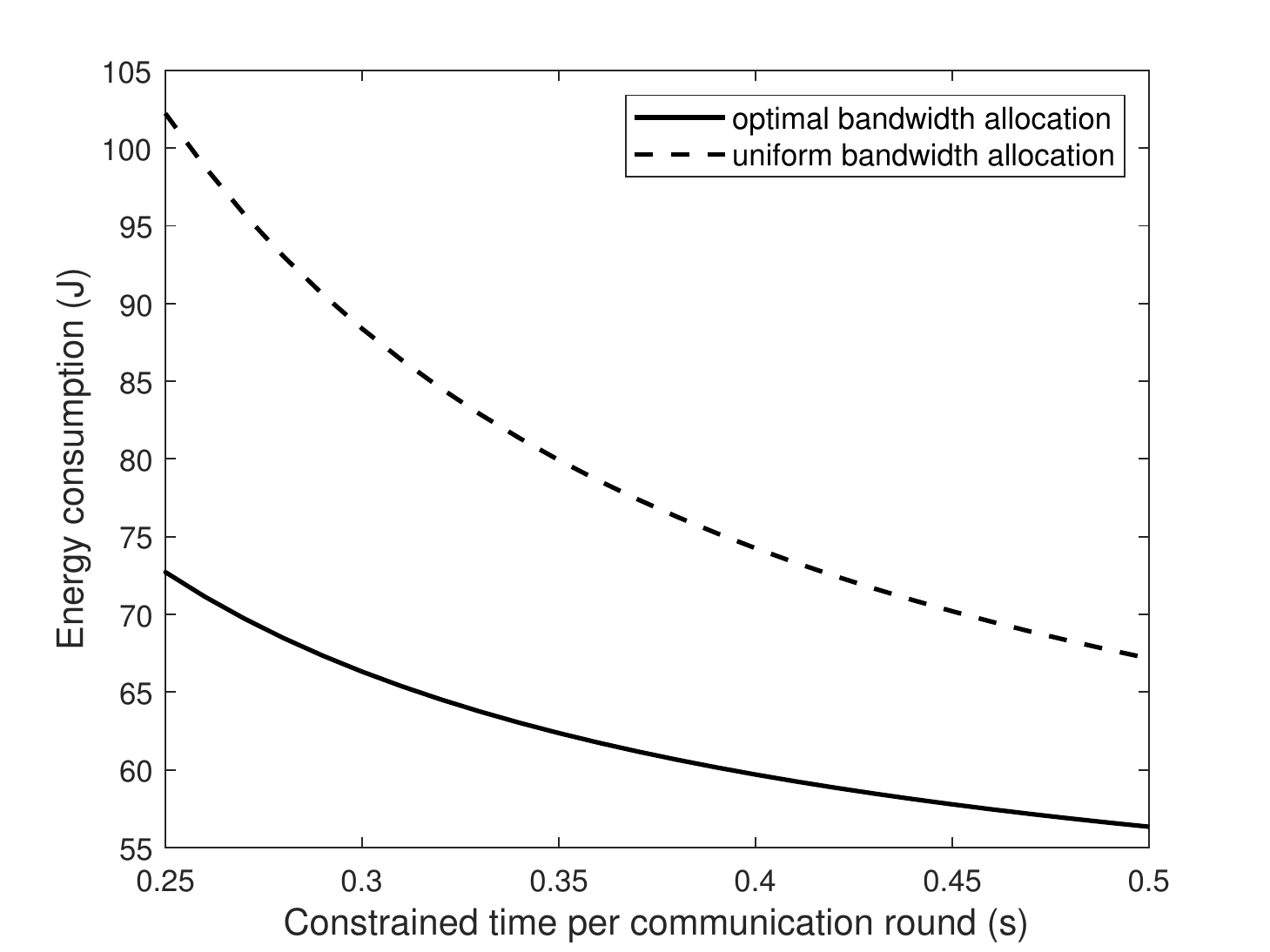}
    \vspace{-5mm}
    \caption{Sum device energy consumption vs. constrained time per communication round in a FEEL system.}
    % \vspace{-10mm}
      \label{Fig:energy_consumption}
\end{figure}

Consider the scenario that all edge-devices are scheduled for model uploading, the performance of the proposed RRM policy is benchmarked against the \emph{uniform bandwidth allocation} policy, which allocates equal bandwidth to edge devices. Particularly, the curves of total energy consumption by edge devices versus the communication round time $T$ are shown in Fig.~\ref{Fig:energy_consumption}. Several observations can be made as follows. First, the total energy consumption reduces as $T$ grows for both cases. This coincides with Lemma~\ref{lemma_non_increasing} that the energy consumption is smaller if the allowed transmission time is larger. Second, it can be found that the proposed optimal policy outperforms the baseline scheme, showing its effectiveness.
\setlength{\textfloatsep}{10pt}
\subsubsection{Energy-and-learning aware scheduling} 
\begin{figure}[t!]
    \centering
    \includegraphics[width=0.7\textwidth]{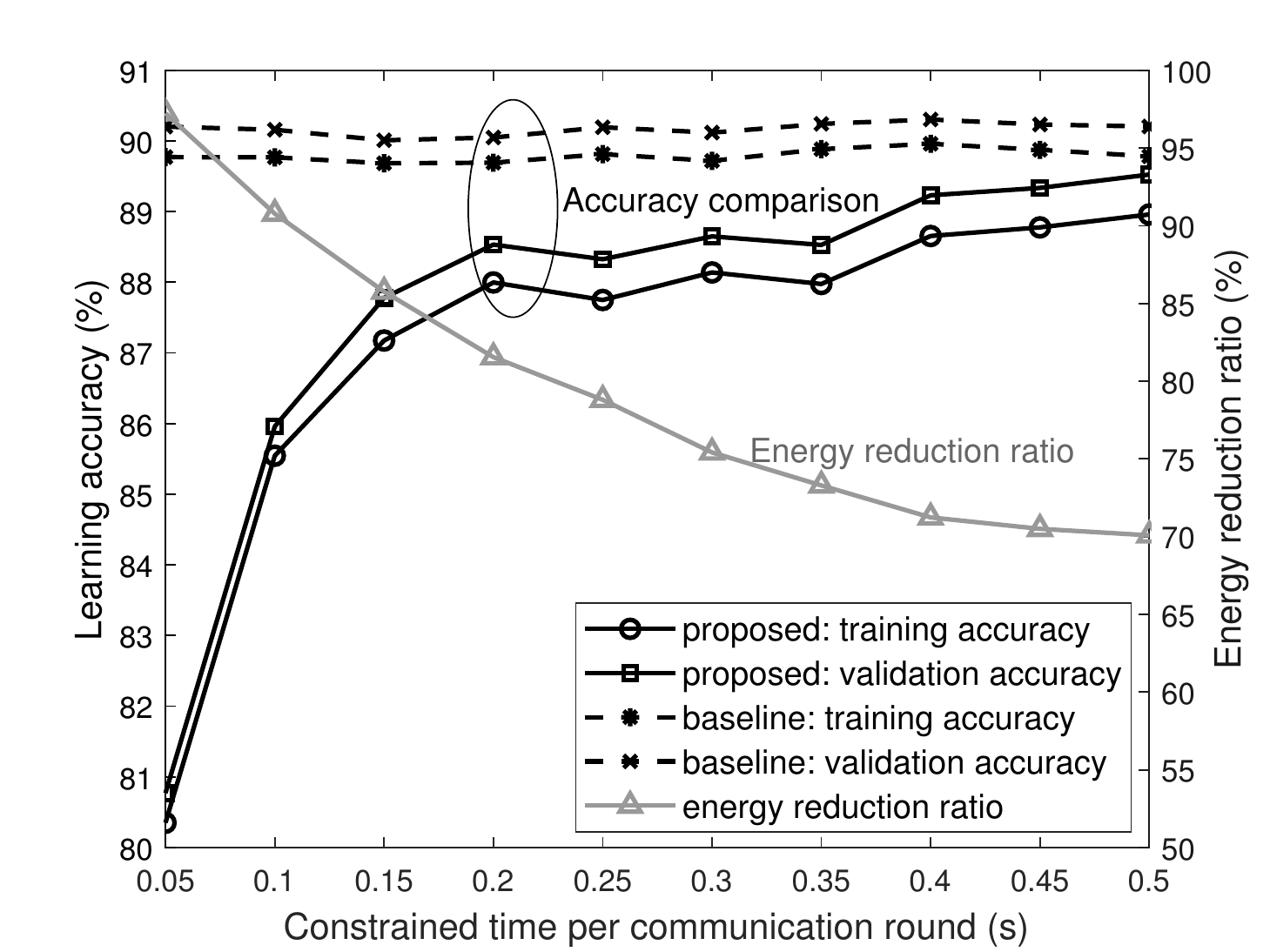}
    \vspace{-5mm}
    % \caption{The solid lines show the expected number of scheduled devices (gray bar/line) and learning accuracy (black solid lines) vs. constrained communication-round time for the proposed energy-and-learning aware scheduling in a FEEL system. The dashed lines show the learning accuracy (black dashed lines) vs. constrained communication-round time for the baseline case which schedules all the devices ($K=50$) for model uploading in the same FEEL system.}
    \caption{The learning accuracy vs. constrained communication-round time for the proposed scheme and the baseline are illustrated by the black solid and dashed lines, respectively. By defining the \emph{energy reduction ratio} as $r = \frac{E_{\sf baseline}-E_{\sf proposed}}{E_{\sf baseline}}\times 100\%$ with $E_{\sf proposed}$ and $E_{\sf baseline}$ denoting the sum energy consumptions of the proposed scheme and the baseline, respectively, the relationship between the energy reduction ratio and constrained communication-round time is shown by the grey line. The implementation details are specified as follows. The total number of  devices is $K=50$ and the communication-round is set to be $10$.}
    % \vspace{-10mm}
      \label{Fig:user_selection}
\end{figure}

Consider the scenario that the communication time is short and the edge server needs to select the edge-devices for uploading. The performance of the proposed Algorithm 1 for joint bandwidth allocation and scheduling is benchmarked against the previous case that all edge-devices are selected. Particularly, the relationship between the average learning accuracy of the federated learning algorithm and the constrained time $T$ is illustrated in Fig.~\ref{Fig:user_selection} given the fixed communication round $10$. Several observations can be made as follows. First, the performance of the baseline is independent of $T$. The reason is that the learning performance only depends on the number of scheduled edge-devices for uploading (i.e. $\sum_{k=1}^K\beta_k$), and this number is fixed for the baseline (i.e. $K=50$). Second, the average learning accuracy of the proposed algorithm is an increasing function of $T$, whose performance approaches the baseline for the large $T$. This is because of the fact that as the allowed transmission time increases, more devices will be scheduled for model uploading, giving rise to the improvement of the learning performance. Furthermore, define the \emph{energy reduction ratio} as $r = \frac{E_{\sf baseline}-E_{\sf proposed}}{E_{\sf baseline}}\times 100\%$, where $E_{\sf proposed}$ and $E_{\sf baseline}$ denote the sum energy consumptions of the proposed scheme and the baseline, respectively. It can be observed that the energy reduction ratio $r$ is a decreasing function of $T$, which approximately ranges from $70\%$ to $98\%$. This is because that as $T$ increases, the scheduled devices in the proposed scheme increases, and thereby the resulting sum energy consumption is larger. This reduces its difference to the sum energy consumption of the baseline, where all devices are scheduled for uploading. 

%our proposed scheduling method reduces sufficiently large sum energy consumption compared with the baseline case. The energy consumption gain, which is defined as the reduction of the proposed sum energy consumption in comparison with the baseline, decreases as $T$ increases. It can be explained from two aspects. 1) For a small $T$, fewer devices are scheduled for uploading in our proposed scheme, leading to much less sum energy consumption when compared to the baseline with all devices for uploading. As $T$ increases, more devices are scheduled, resulting in the difference of device number between the proposed strategy and the baseline diminishing. 2) The proposed optimal bandwidth allocation policy greatly surpasses the baseline, which is verified in earlier simulation, and the effect is more significant for a smaller $T$. 
\setlength{\textfloatsep}{10pt}

%============================================================
\section{Concluding remarks}\label{conclusion}
In this paper, we have proposed energy-efficient RRM (bandwidth allocation and scheduling) for federated edge learning. By adapting to both channel states and computation capacities, the strategies effectively reduce sum device energy consumption while providing a guarantee on learning speed. This work makes the first attempt to explore the direction of energy-efficient RRM for federated edge learning. In the future, this work can be generalized into RRM for the asynchronous model-update scenario. Apart from the channel states and computation capacities, the sparsity of the updates can be also considered while allocating radio resources. Moreover, the effects of energy consumption model for local computing can be further taken into consideration to include the feature of local batch-size adaptation.

%============================================================
\appendix

\subsection{Proof of Theorem \ref{thm_opt_rrm}}\label{proof_opt_tk}
As aforementioned, one can have that $t^{\star}_k = T_k, \forall k$. Next, we prove the optimal bandwidth allocation strategy. Substituting $t_k = T_k$ into $({\bf P1})$, it follows that the original Problem $({\bf P1})$ can be rewritten as
\begin{align}
\begin{split}
    \min_{\gamma_k}&~\sum_{k=1}^K\frac{\gamma_kBT_kN_0}{h_k^2}\left(2^{\frac{\beta_kL}{\gamma_kBT_k}}-1\right)\\
    \text{s.t. }&~~0\leq \gamma_k\leq 1,~k\in\mathcal{K},~~\sum_{k=1}^K\gamma_k=1.
\end{split}
\end{align}

Since the above problem is a convex problem, by introducing Lagrange multipliers $\boldsymbol\mu^{\star} = [\mu^{\star}_1,\mu^{\star}_2,\cdots,\mu^{\star}_K]^T\in\mathbb{R}^{K}$ for the inequality constraints $\boldsymbol\gamma\succeq 0$ with $\boldsymbol\gamma = [\gamma_1,\gamma_2,\cdots\gamma_K]^T$, and a multiplier $\nu^{\star}\in\mathbb{R}$ for the equality constraint $\mathbf{1}^T\boldsymbol\gamma=1$, the KKT conditions can be written as follows
\begin{equation}\label{eqns:KKT}
\begin{gathered}
    \boldsymbol\gamma^{\star}\succeq 0,\quad \mathbf{1}^T\boldsymbol\gamma^{\star}=1,\quad \boldsymbol\mu^{\star}\succeq 0,\quad \mu_k^{\star}\gamma_k^{\star}=0,\quad k\in\mathcal{K}\\
    \frac{BT_kN_0}{h_k^2}\left(2^{\frac{\beta_kL}{\gamma_k^{\star}BT_k}}-\frac{\beta_kL\ln{2}}{\gamma_k^{\star}BT_k}2^{\frac{\beta_kL}{\gamma_k^{\star}BT_k}}-1\right)-\mu_k^{\star}+\nu^{\star}=0,\quad k\in\mathcal{K}.
\end{gathered}
\end{equation}
By solving the above equations, one can have 
\begin{align}\label{eqn:gamma_star}
    \gamma_k^{\star}=\frac{\beta_kL\ln{2}}{BT_k\left[1+\mathcal{W}\left(\frac{h_k^2\nu^{\star}-BT_kN_0}{BT_kN_0e}\right)\right]},
\end{align}
where $\mathcal{W}(\cdot)$ is the Lambert $W$ function, and the Lagrange multiplier value $\nu^{\star}$ is calculated by solving $ \sum\limits_{k=1}^K\frac{\beta_kL\ln{2}}{BT_k\left[1+\mathcal{W}\left(\frac{h_k^2\nu^{\star}-BT_kN_0}{BT_kN_0e}\right)\right]}=1$.
This completes the whole proof.

%------------------------------------------------------------
\subsection{Proof of Corollary~\ref{gamma_Tk}}\label{proof_gamma_Tk}
First, we prove that $\gamma_k^{\star}$ is non-increasing with respect to $T_k$. Denote $x=\frac{h_k^2\nu^{\star}-BT_kN_0}{BT_kN_0e}$, then it follows that $T_k=\frac{h_k^2\nu^{\star}}{\left(x+\frac{1}{e}\right)BN_0e}$. Substituting it to the expression for $\gamma_k^{\star}$, one can have
\begin{align}
    \gamma_k^{\star}=\frac{\beta_{k} L \ln 2}{B T_{k}\left[1+\mathcal{W}\left(\frac{h_{k}^{2} \nu^{\star}-B T_{k} N_{0}}{B T_{k} N_{0} e}\right)\right]}=\frac{ N_{0}e \beta_kL \ln 2}{h_{k}^{2} \nu^{\star}} \frac{x+\frac{1}{e}}{1+\mathcal{W}(x)}.
\end{align}
Further, we denote 
\begin{align}
    y=\frac{x+\frac{1}{e}}{1+\mathcal{W}(x)}=\frac{\mathcal{W}e^{\mathcal{W}(x)}+\frac{1}{e}}{1+\mathcal{W}(x)}.
\end{align}
It is easy to prove that $y$ is non-decreasing with respect to $\mathcal{W}(x)$. Since $\mathcal{W}(x)$ is non-decreasing with respect to $x$ and $x(T_k)$ is non-increasing with respect to $T_k$, it follows that $\gamma_k^{\star}$ is non-increasing with respect to $T_k$.

Next, we prove that $\gamma_k^{\star}$ is non-increasing with respect to $h_k^2$. From $x=\frac{h_k^2\nu^{\star}-BT_kN_0}{BT_kN_0e}$, one can have $h_k^2=\frac{BN_0eT_k}{\nu^{\star}}\left(x+\frac{1}{e}\right)$. Substituting it into the expression for $\gamma_k^{\star}$, it follows that
\begin{align}
     \gamma_k^{\star}=\frac{\beta_{k} L \ln 2}{B T_{k}\left[1+\mathcal{W}\left(\frac{h_{k}^{2} \nu^{\star}-B T_{k} N_{0}}{B T_{k} N_{0} e}\right)\right]}=\frac{\beta_kL\ln{2}}{BT_k}\frac{1}{1+\mathcal{W}(x)}.
\end{align}
Further, we let
\begin{align}
    z=\frac{1}{1+\mathcal{W}(x)}.
\end{align}
It is obvious that $z$ is non-increasing with respect to $\mathcal{W}(x)$. Since $\mathcal{W}(x)$ is non-decreasing with respect to $x$ and $x(h_k^2)$ is non-decreasing with respect to $h_k^2$, we can conclude that $\gamma_k^{\star}$ is non-increasing with respect to $h_k^2$.
This completes the whole proof.

%------------------------------------------------------------
\subsection{Proof of Theorem \ref{thm_opt_user}}\label{proof_opt_betak}

Denote $\boldsymbol\beta=[\beta_1,\beta_2,\cdots,\beta_K]^T\in\mathbb{R}^{K}$ and define the function as follows:
\begin{align}
    J(\boldsymbol\beta)=\sum_{k=1}^K\left[\frac{\gamma_kBT_kN_0}{h_k^2}\left(2^{\frac{\beta_kL}{\gamma_kBT_k}}-1\right)-\lambda \beta_k\right],
\end{align}
then it follows that
\begin{align}
    \frac{\partial J(\boldsymbol\beta)}{\partial \beta_k}=\frac{N_0L\ln{2}}{h_k^2}2^{\frac{\beta_kL}{\gamma_kBT_k}}-\lambda,
\end{align}
Let $\frac{\partial J(\boldsymbol\beta)}{\partial \beta_k}=0$ and one can obtain the following result:
\begin{align}
    \hat{\beta}_k=\frac{\gamma_kBT_k}{L}\log\left({\frac{\lambda h_k^2}{N_0L\ln{2}}}\right).
\end{align}
When considering the constraint, it can be divided into three cases with respect to $\hat{\beta}_k$:
\begin{itemize}
    \item[1)] if $\hat{\beta}_k<0$, then the minimum will be obtained at $\beta_k=0$;
    \item[2)] if $0\leq \hat{\beta}_k\leq 1$, then the minimum will be obtained at $\beta_k=\hat{\beta}_k$;
    \item[3)] if $\hat{\beta}_k>1$, then the minimum will be obtained at $\beta_k=1$.
\end{itemize}
In summary, the optimal point is
\begin{align}
    \beta_k^{\star}=\min\left\{\max\left\{\frac{\gamma_kBT_k}{L}\log\left({\frac{\lambda h_k^2}{N_0L\ln{2}}}\right),0\right\},1\right\}.
\end{align}
This completes the whole proof.
%============================================================
\bibliography{Energy-Efficient_RRM_Extended.bib}
\bibliographystyle{IEEEtran}

\end{document}

%% file: header.tex
\newtheorem{theorem}{Theorem}

\newtheorem{lemma}{Lemma}
\newtheorem{corollary}{Corollary}

\newtheorem{remark}{\bf Remark}
\def\phi{\varphi}

\def\({\left(}
\def\){\right)}

\def\b0{{\mathbf{0}}}

\newcommand{\nn}{\nonumber}

%% file: Energy-Efficient_RRM_Extended.bbl
\begin{thebibliography}{1}
\providecommand{\url}[1]{#1}
\csname url@samestyle\endcsname
\providecommand{\newblock}{\relax}
\providecommand{\bibinfo}[2]{#2}
\providecommand{\BIBentrySTDinterwordspacing}{\spaceskip=0pt\relax}
\providecommand{\BIBentryALTinterwordstretchfactor}{4}
\providecommand{\BIBentryALTinterwordspacing}{\spaceskip=\fontdimen2\font plus
\BIBentryALTinterwordstretchfactor\fontdimen3\font minus
  \fontdimen4\font\relax}
\providecommand{\BIBforeignlanguage}[2]{{%
\expandafter\ifx\csname l@#1\endcsname\relax
\typeout{** WARNING: IEEEtran.bst: No hyphenation pattern has been}%
\typeout{** loaded for the language `#1'. Using the pattern for}%
\typeout{** the default language instead.}%
\else
\language=\csname l@#1\endcsname
\fi
#2}}
\providecommand{\BIBdecl}{\relax}
\BIBdecl

\bibitem{gxzhu_2018_edge_learning}
\BIBentryALTinterwordspacing
G.~Zhu, D.~Liu, Y.~Du, C.~You, J.~Zhang, and K.~Huang, ``Towards an intelligent
  edge: Wireless communication meets machine learning.'' [Online]. Available:
  \url{http://arxiv.org/abs/1809.00343}
\BIBentrySTDinterwordspacing

\bibitem{gxzhu2018FEEL}
\BIBentryALTinterwordspacing
G.~Zhu, Y.~Wang, and K.~Huang, ``Low-latency broadband analog aggregation for
  federated edge learning.'' [Online]. Available:
  \url{http://arxiv.org/abs/1812.11494}
\BIBentrySTDinterwordspacing

\bibitem{deniz2019federated_edge_learning}
\BIBentryALTinterwordspacing
M.~M. Amiri and D.~G{\"{u}}nd{\"{u}}z, ``Machine learning at the wireless edge:
  Distributed stochastic gradient descent over-the-air.'' [Online]. Available:
  \url{http://arxiv.org/abs/1901.00844}
\BIBentrySTDinterwordspacing

\bibitem{sqwang_2018_edge_learning}
S.~Wang, T.~Tuor, T.~Salonidis, K.~K. Leung, C.~Makaya, T.~He, and K.~Chan,
  ``When edge meets learning: Adaptive control for resource-constrained
  distributed machine learning,'' in \emph{{IEEE} Conf. Computer Comm.,
  {INFOCOM}}, pp. 63--71, Honolulu, HI, USA, Apr 16--19 2018.

\bibitem{nishio2018background}
\BIBentryALTinterwordspacing
T.~Nishio and R.~Yonetani, ``Client selection for federated learning with
  heterogeneous resources in mobile edge.'' [Online]. Available:
  \url{http://arxiv.org/abs/1804.08333}
\BIBentrySTDinterwordspacing

\bibitem{yang2018background}
\BIBentryALTinterwordspacing
K.~Yang, T.~Jiang, Y.~Shi, and Z.~Ding, ``Federated learning via over-the-air
  computation.'' [Online]. Available: \url{http://arxiv.org/abs/1812.11750}
\BIBentrySTDinterwordspacing

\bibitem{federatedlearning}
B.~McMahan, E.~Moore, D.~Ramage, S.~Hampson, and B.~A. y~Arcas,
  ``{Communication-Efficient Learning of Deep Networks from Decentralized
  Data},'' in \emph{Proc. of the 20th Intel. Conf. Artificial Intell. and
  Statistics}, vol. 54, pp. 1273--1282, Fort Lauderdale, FL, USA, Apr 20--22
  2017.

\bibitem{mei2017GPUenergy}
X.~Mei, Q.~Wang, and X.~Chu, ``A survey and measurement study of {GPU DVFS} on
  energy conservation,'' \emph{Digital Comm. and Networks}, vol.~3, no.~2, pp.
  89--100, 2017.

\bibitem{bernstein2018sgd}
J.~Bernstein, Y.-X. Wang, K.~Azizzadenesheli, and A.~Anandkumar, ``sign{SGD}:
  Compressed optimisation for non-convex problems,'' in \emph{Proc. of the 35th
  Intl. Conf. Mach. Learning (ICML)}, vol. 80, pp. 560--569, Stockholmsmässan,
  Stockholm Sweden, Jul 10--15 2018.

\end{thebibliography}
